\documentclass{IEEEtran}
\usepackage[utf8]{inputenc}

\usepackage{authblk}

\newcommand{\name}{BVOT\xspace}
\newcommand{\enc}{\mathcal{E}}

\newtheorem{definition}{Definition}
\newtheorem{theorem}{Theorem}

\usepackage{amssymb,amsfonts}
\usepackage{paralist}
\usepackage{times,framed}
\usepackage[noend]{algpseudocode}
\usepackage{algorithm}
\usepackage{verbatim}
\usepackage{flushend}
\usepackage{lipsum}
\usepackage{graphics}
\usepackage{epsfig}
\usepackage{amsmath}
\usepackage{xcolor}
\usepackage{lipsum}
\usepackage{xspace}
\usepackage{soul}
\usepackage{caption}
\usepackage{subcaption}

\title{BVOT: Self-Tallying Boardroom Voting with Oblivious Transfer}
\date{October 2020}
\author[ ]{Farid Javani}
\author[ ]{Alan T. Sherman}
\affil[ ]{Cyber Defense Lab, University of Maryland, Baltimore County (UMBC), Maryland, USA}

\affil[ ]{ {\texttt{\{javani1,sherman\}@umbc.edu}}}

\begin{document}
\maketitle

\begin{abstract}
A {\it boardroom election} is an election with a small number of voters
carried out with public communications.
We present BVOT, a self-tallying boardroom voting protocol with ballot secrecy, 
fairness (no tally information is available before the polls close), and 
dispute-freeness (voters can observe that all voters correctly followed the protocol). 

BVOT works by using a multiparty threshold homomorphic encryption system in
which each candidate is associated with a masked unique prime.
Each voter engages in an oblivious transfer with an untrusted {\it distributor}: 
the voter selects the index of a prime associated with a candidate and receives the selected prime 
in masked form. The voter then casts their vote by encrypting their masked prime
and broadcasting it to everyone.
The distributor does not learn the voter's choice, and no one learns
the mapping between primes and candidates until the audit phase.
By hiding the mapping between primes and candidates, BVOT provides voters
with insufficient information to carry out effective cheating.
The threshold feature prevents anyone from computing any partial 
tally---until everyone has voted.
Multiplying all votes, their decryption shares, and the unmasking factor
yields a product of the primes each raised to the number of votes received.

In contrast to some existing boardroom voting protocols, 
BVOT does not rely on any zero-knowledge proof;
instead, it uses oblivious transfer to assure ballot secrecy and correct vote casting. 
Also, BVOT can handle multiple candidates in one election.
BVOT prevents cheating by hiding crucial information:
an attempt to increase the tally of one candidate 
might increase the tally of another candidate.
After all votes are cast, any party can tally the votes. 

\textbf{keywords:} Applied cryptography, boardroom voting, election systems, oblivious transfer.

\end{abstract}

\section{Introduction}\label{sec:into}

Many of the proposed high-integrity election systems 
depend on an election authority to administer elaborate procedures
or require voters to carry out complex steps, such as 
executing and checking {\it zero-knowledge proofs (ZKPs)}.
We propose a new boardroom voting protocol, \name, 
which is self-tallying (any voter can tally the votes)
and is based on {\it oblivious transfer (OT)}.

We focus on electronic {boardroom voting}, such as that
carried out at a meeting of shareholders or a board of directors.  
A {\it traditional boardroom election} is an election that takes place in a single room,
where the voters can see and hear each other~\cite{blanchard2020boardroom}. 
The election is conducted by an untrusted party, who could be a voter, also present in the room. 
We address an {\it electronic version of boardroom elections} 
in which a small number of voters 
participate in person or remotely through web browsers or applications, in the absence of
a substantial election authority.
The small scale of boardroom elections permits the use of protocols and cryptographic primitives 
that might be impractical at large scale.
Examples of boardroom voting include
\cite{kiayias2002self,groth_2004,hao_2010,k12f}.
By contrast, many other voting systems require substantial
election authorities (e.g.,~\cite{fujioka1992practical,Adida2008HeliosWO,clarkson2008civitas,chaum2009scan,remotegrity13,hao2014every,shahandashti2016dre}).

We seek an electronic boardroom voting system that provides each of the 
following properties~\cite{hao_2010,kiayias2002self}: 

\begin{itemize}

\item \textbf{Fairness.} None of the voters can learn a complete or partial tally of the votes before casting their vote.

\item \textbf{Dispute-freeness.} Each voter can observe if the other voters have carried out the the protocol correctly.

\item \textbf{Perfect ballot secrecy.} How each voter voted remains secret during and after the election, and a partial tally of the votes of any subset of voters is possible only with the collaboration of all of the other voters.  

\item \textbf{Self-tallying.} Any voter can compute the tally.

\end{itemize}

The main novel feature of \name is its use of OT 
to provide perfect ballot secrecy and ensure correct vote casting.
Doing so avoids the need for voters to carry out or verify complex ZKPs.
Given that Kilian~\cite{kil88} proved that oblivious transfer is complete for two-party secure computations, it is intriguing to explore applications of OT in voting.
Nurmi et al.~\cite{nurmi91} also used OT in an election system, but
only to distribute credentials to voters.
\name can handle multiple candidates in one election.
By contrast, \cite{kiayias2002self,groth_2004,hao_2010,k12f}
require extensions with additional performance costs to
handle multiple candidates.

\begin{figure}[ht!]

\begin{subfigure}{\columnwidth}
\centering
\fbox{
\includegraphics[width=.6\columnwidth]{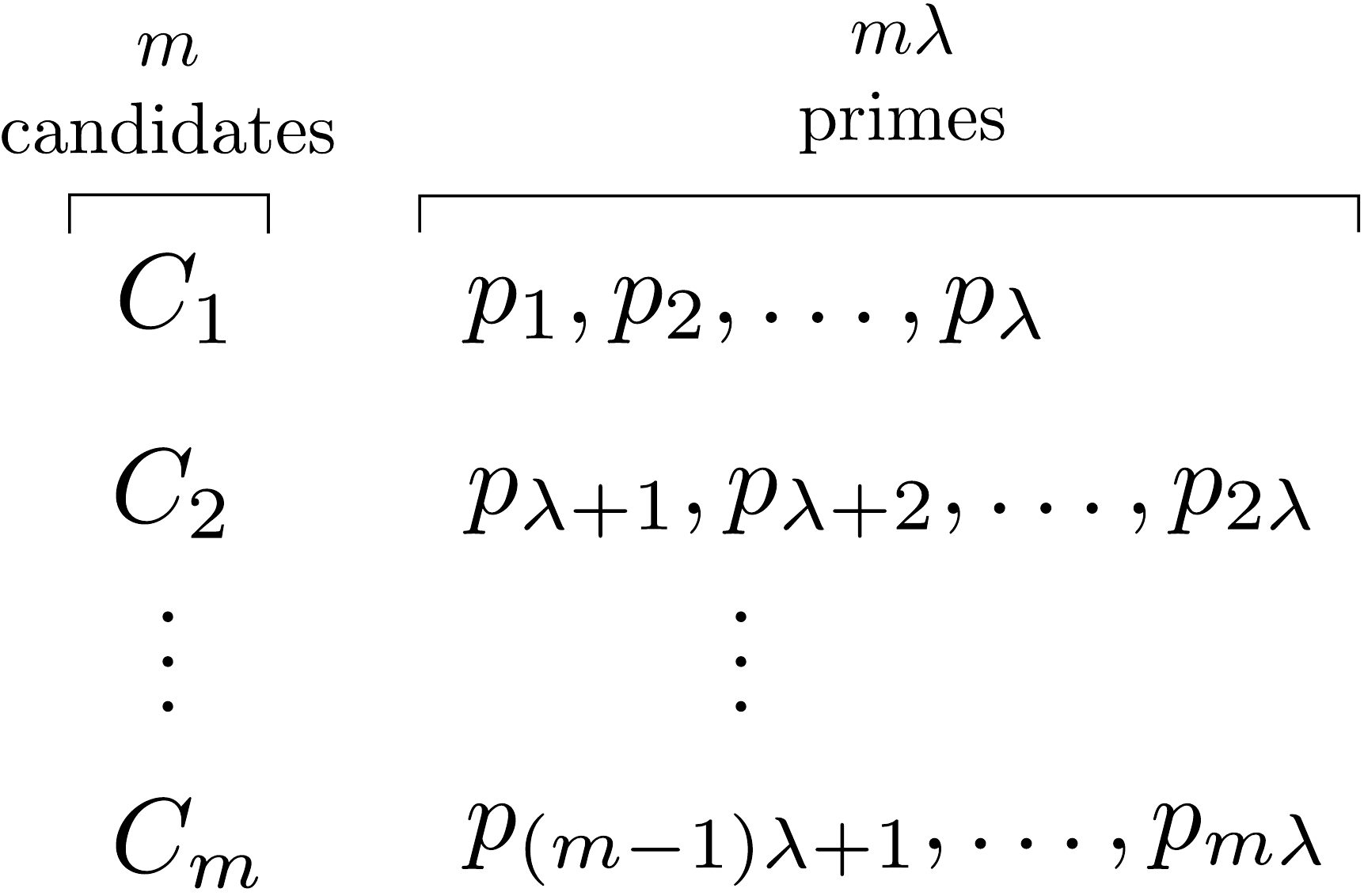}
}
\caption{Prime-to-candidate mapping. Each prime is associated with a candidate.}

\end{subfigure}
\vspace{2mm}

\begin{subfigure}{\columnwidth}
\centering
\fbox{
\includegraphics[width=.6\columnwidth]{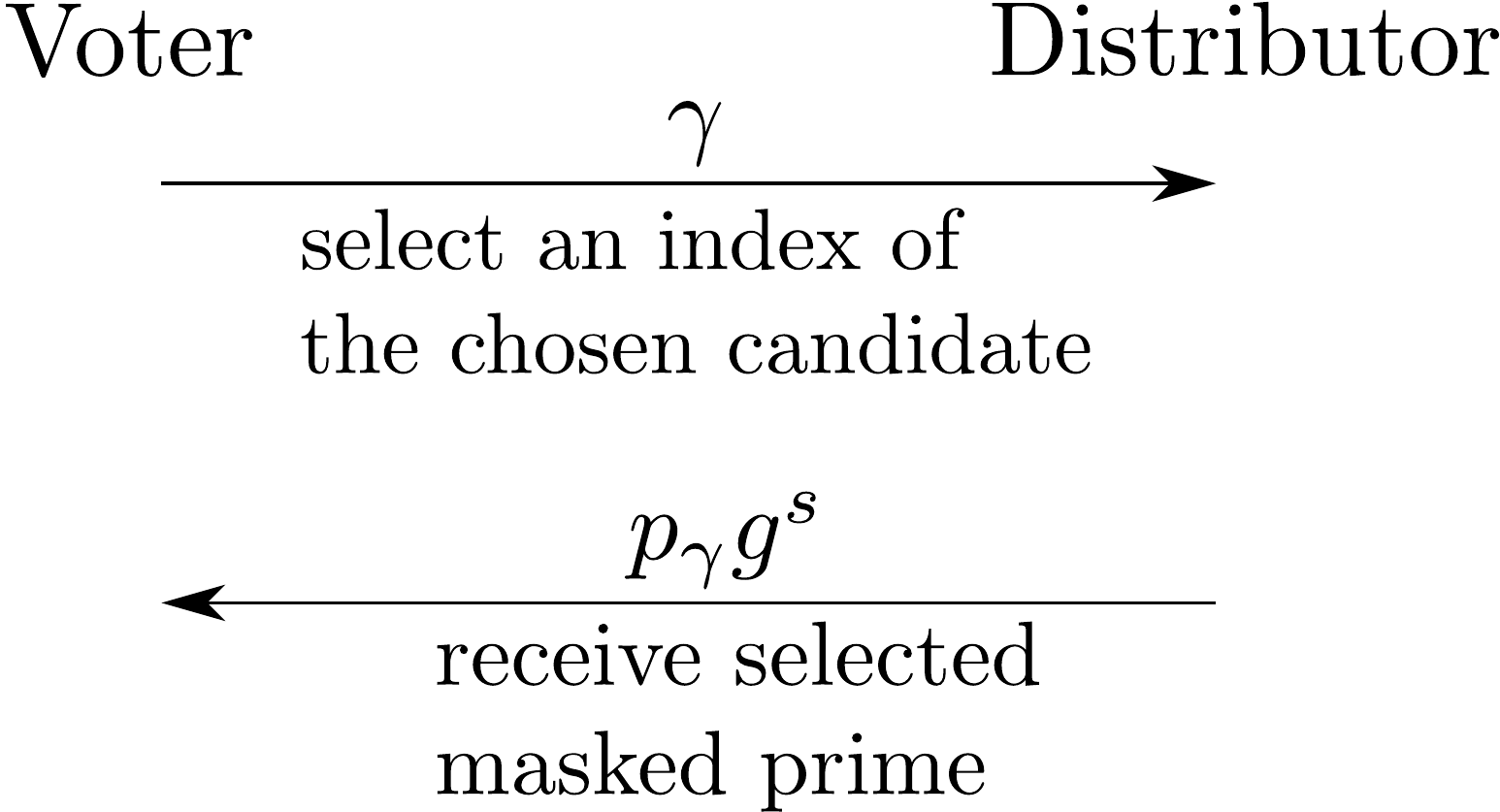}
}
\caption{Oblivious transfer between voter and distributor.}

\end{subfigure}

\caption{In BVOT, there are $\lambda$ primes associated with each of the $m$ candidates.
    Each voter selects the index $\gamma$ of an unknown prime $p_{\gamma}$ associated with their chosen candidate and,
    through an $\textrm{OT}^{\lambda m}_1$ with an untrusted distributor, receives the selected prime 
    in masked form $p_{\gamma}g^s$. The distributor does not learn the voter's choice, 
    and no one learns the mapping between primes and candidates until the audit phase.}
    \label{fig:BVOT}
    
\end{figure}

As shown in Figure~\ref{fig:BVOT},
\name works by using a multiparty threshold homomorphic encryption system in
which each candidate is associated with a masked unique prime.
Each voter engages in an OT with an 
untrusted {\it distributor}: 
the voter selects the index of a prime associated with a candidate and receives the selected prime 
in masked form.  
The distributor is untrusted both for privacy and integrity.
The voter then casts their vote by encrypting their masked prime
and broadcasting it to everyone.
The distributor does not learn the voter's choice, and no one learns
the mapping between primes and candidates until the audit phase.

By hiding the mapping between primes and candidates, \name provides voters
with insufficient information to carry out effective cheating.
The threshold feature prevents anyone from computing any partial 
tally---until everyone has voted.
Multiplying all votes, their decryption shares, and the unblinding factor
yields a product of the primes each raised to the number of votes received. 
The small size of a boardroom election enables everyone to factor this product.
BVOT's novel use of primes to represent ballot choices enables BVOT
to handle multiple candidates without extensions.

Our contribution is a remote self-tallying boardroom voting protocol based on OT that is
fair and dispute free and enjoys perfect ballot secrecy.

\section{Oblivious Transfer}\label{sec:Background}

We briefly review our main building block---{\it oblivious transer (OT)}---including selected OT protocols and their security and efficiency.

First introduced by Rabin~\cite{Rabin81}, an OT protocol enables a receiver to receive a piece of information from a sequence of pieces of information from a sender, while hiding the selection of information from the sender and hiding the rest of the information from the receiver. Formally, in  1-out-of-2 OT, denoted $\textrm{OT}^2_1$, the sender has two strings $s_0,s_1$ and transfers $s_b$ to the receiver, where 
the receiver selects $b\in \{ 0,1 \}$ and the following two conditions hold: 
(1)~the sender does not know the value of $b$, and 
(2)~the receiver does not learn anything about $s_{1-b}$.

We will use a generalization 1-out-of-$n$ OT, denoted $\textrm{OT}^n_1$: the sender has $n$ strings and transfers one string to the receiver, without knowing which string it transferred, and the receiver does not learn anything about the other $n-1$ strings.

An \textit{ideal implementation} of OT uses a trusted third party: after obtaining the strings from the sender, and the index choice from the receiver, the trusted party sends the chosen string to the receiver. 

OT can be implemented using public-key cryptography without a trusted third party. For example, 
$\textrm{OT}^2_1$ can be implemented as follows:
the receiver creates two random public keys but knows the private key corresponding to only one of them. The receiver sends the two public keys to the sender. The sender encrypts each string with a different public key and sends the resulting ciphertexts to the receiver. 
Because the receiver knows the private key corresponding to only one of the public keys, 
the receiver will be able to decipher only one of the strings,
and the receiver will learn nothing about the other string.

Implementing OT with public-key operations, however, is computationally expensive. 
Seeking faster implementations, researchers have explored the possibility of implementing OT using symmetric-key cryptography, but Impagliazzo and Rudich~\cite{impagliazzo1989limits} showed that it is unlikely to find black-box constructions of OT using one-way functions.

Seeking greater efficiency,
Bellare and Micali~\cite{Bellare89} created an $\textrm{OT}^2_1$ that requires two rounds. 
Naor and Pinkas~\cite{Naor01} reduced the number of exponentiations during run-time in Bellare and Micali from two to one on the sender's side. 
They also extended $\textrm{OT}^2_1$ to $\textrm{OT}^n_1$. 
In this $\textrm{OT}^n_1$ technique, the sender performs $n$ exponentiations in the initialization step, and uses the resulting values for all subsequent transfers.

Noar and Pinkas~\cite{naor2005compsec} showed how to extend an $\textrm{OT}^2_1$ protocol to an $\textrm{OT}^n_1$ protocol---with $O(n\log n)$ calls to $\textrm{OT}^2_1$---that provides sender
and receiver security computationally, if the  underlying $\textrm{OT}^2_1$ provides sender and receiver security (see Section~\ref{sec:security} for definitions). 
Among the most efficient OT protocols that are secure against active adversaries (including possibly
sender and/or receiver)
are~\cite{peikert2008framework,chou2015simplest,asharov2017more}. 
In \name, vote correctness and ballot secrecy rely on the security of OT, so it is crucial that the OT is secure against an active adversary. 

Because there are few voters in \name, the performance of the underlying OT is tolerable:
\name uses one $\textrm{OT}^n_1$ for each voter, and these $\textrm{OT}^n_1$s can be executed in parallel. 
Chou and Orlandi~\cite{chou2015simplest} computed more than $10,000$ $\textrm{OT}^2_1$s per second 
using one thread of an Intel Core i7-3537U processor.
Even with the overhead of building each $\textrm{OT}^n_1$ from $\textrm{OT}^2_1$s
using Noar and Pinkas's technique,
the $\textrm{OT}^n_1$s in \name can be executed sufficiently quickly.

\section{Previous and Related Work} 

We briefly review selected self-tallying boardroom voting protocols that provide ballot secrecy. 
We also review selected protocols that use OT as a primary building block.

\subsection{Self-Tallying Boardroom Voting Protocols}

Kiayias and Yung~\cite{kiayias2002self} proposed the first self-tallying boardroom voting protocol with perfect ballot secrecy. 
Groth~\cite{groth_2004} simplified the protocol and reduced its computational complexity, preserving the same security properties. 
Hao {et al.}~\cite{hao_2010} proposed a similar self-tallying, dispute-free protocol with perfect ballot secrecy that needs only two rounds of communication.

In the protocols of Hao {et al.} and Groth, for each $1 \leq i \leq n$, voter $V_i$ 
chooses a vote $v_i \in \{0,1\}$ ($1$ for ``yes'' and $0$ for ``no'') and 
computes $g^{v_i}$, where $g$ is a generator of a group in which the Diffie-Hellman assumption holds \cite{stinson2018cryptography}. 
Each voter publishes $g^{v_i} A_i$ as their masked vote, for some secret masking value $A_i$.
Hao generates $A_i$ differently than does Groth.
In each protocol, the $A_i$'s are chosen so that $\prod_{i=1}^{n} A_i =1$. 
Therefore, $\prod_{i=1}^{n} g^{v_i}A_i=g^{\sum v_i}$. 
The number of ``yes'' votes is
$\sum v_i$.  Because this sum is small, it can be easily calculated given $g^{\sum v_i}$,
even assuming the discrete logarithm problem is hard.

In Hao {et al.} and Groth, using a ZKP, each voter proves that
they voted correctly by proving $v_i \in \{0,1\}$ (and not, for example, $v_i=2$). 
The correctness of their protocols depends on this ZKP.
By contrast, \name does not require such proofs.

The main difference between the protocols of Hao {et al.} and Groth 
is how they compute $\prod_{i=1}^{n} g^{v_i}A_i$. 
In Hao {et al.}, voters publish their masked votes in one round, enabling
this product to be computed immediately.
In Groth, voters compute the product sequentially.
Consequently, Hao {et al.}'s protocol requires two rounds of communication, whereas Groth 
requires $n+1$ rounds. 
Our protocol is similar to Hao {et al.}'s and requires only five
rounds of communication.

Szepieniec and Bart~\cite{s15n} proposed a protocol 
similar to Kiayias and Yung that also provides fairness. 
Giustolisi {et al.}~\cite{giustolisi2016possibility} showed how to reuse the 
key-sharing round in Hao {et al.}, reducing the number of rounds to one in all subsequent elections. 
Adding a commitment round to Hao {et al.}, Khader {et al.}~\cite{k12f} achieve fairness, 
and, by introducing a recovery round, they achieve robustness (Section~\ref{sec:robust}).

In each of these protocols, 
because each vote is 0 or 1, the protocol must be executed once for each candidate.  
Cramer {et al.}~\cite{cramer97} proposed a technique for handling multiple candidates 
in one protocol execution by using
independent generators for the underlying group, one per candidate.
Their technique complicates the ZKPs.
By contrast, without any ZKPs, \name supports multiple candidates in one protocol execution by
using a separate small prime integer for each candidate.

\subsection{Applications of Oblivious Transfer} 

OT is a powerful primitive that can be used alone to implement any secure two-party~\cite{kil88} and multiparty~\cite{crepeau1995committed} secure computation. We briefly point out three examples.

Nurmi {et al.}~\cite{nurmi91} used $\textrm{OT}^n_1$ to enable a trusted election authority
to distribute a credential to each of the $n$ voters, such that the election authority
does not know the credential of any voter. Each voter uses their credential to cast their ballot.
OT prevents the election authority from linking voters and their ballots. 
By contract, \name does not depend on a trusted third party, and 
\name uses OT to prevent cheating and to hide primes associated with candidates.

Even {et al.} used OT to sign contracts~\cite{even85}.
Two parties use $\textrm{OT}^2_1$ to exchange secrets, where
knowledge of the other's secret implies their commitment to the contract. 
Here, OT guarantees: 
each party sends its secrets correctly, and 
both parties simultaneously exchange their secrets.

Fagin {et al.}~\cite{fagin96} used OT to enable two parties to compare their secrets without 
revealing them
(e.g., a user wants to prove their identity using a password but does not trust the medium).

\section{The BVOT Voting System}\label{sec:Protocol}

We explain the \name self-tallying boardroom voting system by describing 
the election organization, our adversarial model, 
the underlying homomorphic encryption system, 
the voting protocol, and a small example.

\subsection{Election Organization}

There is a set of untrusted voters, and one of them is designated as the {\it distributor}.
The distributor generates certain values and uses OT to distribute them to the other voters.
After all votes are cast and published, anyone---including voters and non-voters---can tally the votes.
Instead of using a bulletin board to post information, 
for simplicity, BVOT relies on broadcasting messages.

\subsection{Adversarial Model}
\label{sec:adversary}

Voters communicate with each other from trusted machines (running a trusted app) 
over an authenticated channel.
A covert polynomial-time adversary (possibly a voter) 
listens to all communications.

The adversary and voters are cautious: they will follow the protocol 
and they do not want to be caught misbehaving, 
but they may try to cheat in favor of some candidate or to learn how others voted. 
Subject to this constraint, the adversary can behave actively, including as sender or receiver
of an OT.

The adversary cannot break standard cryptographic functions.
We assume that the adversary does not try to sabotage or delay the election.
In particular, in the spirit of boardroom elections, we assume that none of the voters
will intentionally lie (e.g., make a false claim of malfeasance) for the purpose of discrediting the election.

\subsection{The Homomorphic Encryption System} \label{sec:prelim}

\name uses a multiparty threshold encryption system based on ElGamal, similar to that 
described by Benaloh~\cite{benaloh2006}. 
Voters agree on a 
multiplicative group $G=\mathbb{Z}_q^*$ of order $q-1$, with generator $g$, 
for which group the Diffie-Hellman problem is hard~\cite{stinson2018cryptography}.
Here, $q$ is a large prime such that $q-1$ has at least one large factor.

\paragraph{Key Generation.}
Each voter $V_i$ chooses a private key\linebreak $d_i \in_R G$ 
at random and sends $g^{d_i}$ to all other voters. The product
$e=\prod_{i=1}^{n} g^{d_i}$ is the {\it public key of the set of $n$ voters}.
These keys can be used for multiple elections.
\paragraph{Encryption.} To encrypt any message $M$ using key $e$, voter
$V_j$ computes  $\enc_{V_j}(M)=(g^{x_j},M e^{x_j})$, 
where $x_j \in_R G$ is a value chosen randomly by $V_j$.

\paragraph{Decryption.}
For ciphertext $(g^{x_j},M  e^{x_j})$, the {\it decryption share} of voter $V_i$ 
is $ \left( g^{x_j} \right) ^{-d_i} $. 
To decipher this ciphertext, multiply $M e^{x_j}$ by 
all decryption shares for this ciphertext:

\begin{equation} \label{eq1}
\begin{split}
Me^{x_j}\prod_{i=1}^{n}  \left( g^{x_j} \right) ^{-d_i} & = M\prod_{i=1}^{n} \left( g^{{d_i}} \right)^{x_j}\prod_{i=1}^{n} \left( g^{x_j} \right) ^{-d_i} \\
 & = M g^0 = M.
\end{split}
\end{equation}

\subsection{Boardroom Voting Protocol}\label{sec:vote-prt} 

We describe the four steps of the voting protocol in detail, in which $n$ voters
choose among $m$ candidates.  All cryptographic operations take place 
in a multiplicative group $G$, as defined in Section~\ref{sec:prelim}.
For each positive integer $k$, let\linebreak
${\mathbb Z}_k = \{ 0, 1, 2, \ldots, k-1 \}$.

\begin{enumerate}

\item[Step 1.] \textit{Election Setup}

\begin{enumerate}

\item The voters select a distributor $D$.

\item The voters agree on a parameter $\lambda$, which denotes the number of primes to represent each candidate
(having multiple primes per candidate mitigates certain attacks by creating uncertainty which prime a voter selected).

\item $D$ chooses any $\lambda m$ primes $p_1,p_2,\dots p_{\lambda m}$ such that
$b^n<q$, where $b = \max(p_1,p_2,\dots, p_{\lambda m})$.
Hence, the product of any $n$ of the primes is less than $q$.

\item Using a standard numbering system, $D$ associates 
the index of each of the primes with some candidate.
Let $\mathcal{M} : {\mathbb Z}_{\lambda m} \rightarrow {\mathbb Z}_m$ denote this mapping.
In particular, primes $p_1, p_2,  \ldots, p_\lambda $ are associated with Candidate~1;
primes $p_{\lambda + 1}, p_{\lambda + 2},  \ldots, p_{2\lambda} $ are associated with Candidate~2;
and so forth. 

\item $D$ selects a mask $s \in_R G$ at random. $D$ commits to $\mathcal{M}$ and $s$  
(e.g., by publishing a cryptographic hash of these values).

\item $D$ masks each prime $p$ as $p g^{s}$.

\end{enumerate}
\end{enumerate}

Masking the primes mitigates certain attacks involving substituting one prime for another, in hopes of favoring some candidate: the adversary does not know which prime would affect which candidate.
Furthermore, the distributor knows the mapping but not which primes are used in the votes, 
in part because there are multiple primes per candidate.

As an example of an election setup, consider an election with 128 voters held using
a group with $q$ having 2048 bits. If $b<2^{16}$, then there would be 6542 primes
available to represent the candidates.  In this example, 
$\lambda$ can be chosen such that $\lambda m > 128$ (see Section~\ref{lambda}).

\begin{enumerate} 
\item[Step 2.] \textit{Vote Selection}

\begin{enumerate}

\item To select their candidate, each voter engages in an $\textrm{OT}^{\lambda m}_1$ 
with $D$ as the sender using the $\lambda m$ masked primes.
The candidate selects an index of a prime associated with their candidate,
as given by $\mathcal{M}$. The candidate receives the corresponding masked prime.

\end{enumerate}

As a result of the OT, $D$ does not learn the voter's selection.

\item[Step 3.] \textit{Vote Casting} 

\begin{enumerate}

\item Each non-distributor voter $V_i$ publishes the encryption 
of their chosen masked prime $p_{r_i} g^{s}$,
encrypted with the public key $e$ of the set of voters:
$\enc_{V_i}(p_{r_i} g^{s})=(g^{x_i},(p_{r_i} g^{s})  e^{x_i})$, where
$x_i \in_R G$ is chosen at random.

\item The distributor $D = V_\delta$ publishes the encryption of their unmasked prime:
$\enc_{V_\delta}(p_{r_\delta})=(g^{x_\delta},p_{r_\delta} e^{x_\delta}).$

\end{enumerate}

Encryption prevents anyone from learning the chosen primes, 
even after $s$ is eventually revealed.  

\end{enumerate}

\begin{enumerate}

\item[Step 4.] \textit{Vote Tallying}  

\begin{enumerate}

\item After all votes are cast, $D$ broadcasts
$\left( \prod_{k=1}^{n} g^{x_k}\right)^{-d_\delta}$, 
which is its decryption share 
of the product of the $g^k$ values from the 
encrypted votes from all other\linebreak voters.
Then, each other voter $V_i$ broadcasts\linebreak
$\left( \prod_{k=1}^{n} g^{x_k}\right)^{-d_i}$,
which is its decryption share of this product.

\item $D$ publishes $\mathcal{M}$ (defined in terms of masked primes).
From this value, each voter verifies that they received a masked prime that
corresponds to their chosen candidate.  
This step gives each voter an opportunity to file an allegation that
they received the wrong masked prime---before the voter learns the tally.

\item $D$ publishes $g^{-(n-1)s}$ and $s$. 

\item Anyone can tally the votes by calculating and factoring the product $P$ of 
all encrypted votes, the encryption shares, and $g^{-(n-1)s}$.
This product is the product of the selected candidate primes, each raised to
the number of votes for that prime.  That is, 

\begin{equation}
\begin{split}
P = & \left(( p_{r_\delta}e^{x_\delta}) \prod_{\substack{{i=1} \\ {i\neq \delta}}}^{n} (p_{r_i} g^{s}) e^{x_i}  \right)     \\
&  g^{-(n-1)s} \left( \prod_{k=1}^{n} g^{x_k} \right) ^{-d_\delta}  \prod_{\substack{{i=1} \\ {i\neq \delta}}}^{n} \left( \prod_{k=1}^{n} g^{x_k} \right) ^{-d_i} 
\\
= & p_{r_1} p_{r_2} \dots p_{r_n}\\
= &  p_1^{a_1} p_2^{a_2} \dots p_{\lambda m}^{a_{\lambda m}}.
\end{split}
\end{equation}

\end{enumerate}

The sum of the votes is $a_1+a_2+\dots+a_{\lambda m}=n$, and 
no candidate prime can be selected more than $n$ times:
$0\leq a_1,a_2,\dots, a_{\lambda m}\leq n$. 
Because boardroom elections involve a small number of voters, 
the primes $p_1, p_2, \dots, p_{\lambda m}$ and 
their powers are small.  Therefore, $P$ can be easily factored.

\end{enumerate}

\subsection{An Example with Three Candidates and Four Voters} 

We illustrate BVOT in an election in which four voters \linebreak
$(V_1, V_2, V_3, V_4)$
select among three candidates 1,2,3.  Assume $V_4$ is the distributor $D$.

For this example, we will represent each candidate by $\lambda=3$ distinct primes 
(see Section~\ref{lambda}).
Let these distinct primes be $p_1,p_2, \dots, p_9$. Distributor
$D$ randomly partitions the list of primes to associate three primes with each candidate, 
cryptographically commits to this association, but does not
immediately reveal the association.
$D$ selects a mask $s \in_R G$ at random and masks
each prime $p$ by computing $p g^{s}$.

Using the following standard numbering system, Candidate~1 is associated with indices 1,2,3;
Candidate~2 is associated with indices 4,5,6;
and Candidate~3 is associated with indices 7,8,9.

Using the standard numbering system described above, each voter selects a candidate by engaging in 
an $\textrm{OT}^9_1$ with $D$ to receive one of the nine masked primes.
For example, to select Candidate~2,
the voter would select index 4, 5, or 6 in the OT.  As a result of the OT, the voter
receives the masked prime corresponding to their chosen index.
$D$ does not learn the selected index, and the voter learns only the selected
masked prime.

For each $1 \leq i \leq 3$, let $p_{r_i}g^s$ denote the masked prime received by $V_i$.
Each voter encrypts their masked prime with the public key $e$ of the set of voters.
Each voter publishes the resulting ciphertext by sending it to the other voters.
For example, $V_1$ publishes $ \enc_{V_1}(p_{r_1} g^{s})=(g^{x_1},(p_{r_1} g^{s} )e^{x_1})$. 
Similarly, $D$ publishes the encryption of their selected unmasked prime $p_{r_4}$.

After all encrypted votes are published, $D$ publishes\linebreak
$(\prod_{k=1}^{4} g^{x_k})^{-d_4}$, which is
its decryption share of the product of the
$g^k$ values from
the encrypted votes from the other voters.
Next, each of the other voters publishes 
their decryption share of this product.
For example, $V_1$ publishes $(\prod_{k=1}^{4} g^{x_k} )^{-d_1} $. 

Finally, $D$ publishes the prime-to-candidate mapping, $g^{-3s}$, and $s$.
From the published values, anyone can compute the tally by 
decrypting and unmasking the selected masked primes and computing their product.
Specifically, the product, $P$, 
of all published encrypted masked primes, decryption shares, and $g^{-3s}$ yields
the tally encoded as a product of the nine candidate primes, each raised to the number
of votes for that prime.  To wit, 

\begin{equation}
\begin{split}
P = & ((p_{r_1} g^{s}) e^{x_1})((p_{r_2} g^{s}) e^{x_2})((p_{r_3} g^{s})  e^{x_3})(p_{r_4}  e^{x_4})\\
&
(\prod_{k=1}^{4} g^{x_k})^{-d_1}
(\prod_{k=1}^{4} g^{x_k})^{-d_2}
(\prod_{k=1}^{4} g^{x_k})^{-d_3}\\
&
(\prod_{k=1}^{4} g^{x_k})^{-d_4}
(g^{-3s})\\
=&  p_{r_1} p_{r_2} \dots p_{r_4}
=   { p_1^{a_1}} {p_2^{a_2}} \dots {p_9^{a_9}},
\end{split}
\end{equation}

\noindent
where, for each $1 \leq i \leq 9$, $a_i$ is the number of votes for prime $p_i$.
Thus, $a_1+a_2+\dots+a_9=4$ is the sum of the votes.  In particular, if 
the primes for Candidate~1 were $p_3, p_4, p_9$, then 
Candidate~1 would have received $a_3 + a_4 + a_9$ votes.

\section{Communication and Performance Analysis} \label{sec:perf}

We analyze and compare the computational and communication complexity of \name 
with that of selected other self-tallying boardroom voting protocols.
Our analysis considers one election with
$n$ voters and $m$ candidates, using security parameter $\lambda$.

The key-generation step requires one exponentiation per voter. 
Each voter performs one homomorphic encryption and computes the product of $n$ decryption shares;
each of these steps requires one exponentiation.
\name performs $n-1$ parallel sessions of $\textrm{OT}^{\lambda m}_1$.

The distributor makes one commitment to the mapping $\mathcal{M}$.
In addition to their role as a voter, the distributor performs one more exponentiation to compute $g^{-(n-1)s}$. 

\name requires five broadcast rounds, three of which are performed solely by the distributor: 
(1)~Each voter broadcasts their public keys.
(2)~Each voter broadcasts their encrypted vote.
(3a)~The vote distributor broadcasts its decryption share.
(3b)~Each voter broadcasts their decryption share.
(4a)~The vote distributor broadcasts $\mathcal{M}$.
(4b)~The vote distributor broadcasts $g^{-(n-1)s}$ and $s$.

Table~\ref{tab:compare-perf} compares the computational and communication complexity of \name to that
of four other selected self-tallying boardroom voting protocols.

\setlength{\tabcolsep}{2pt}

\begin{table}[ht]
\caption{Communication and performance of selected self-tallying boardroom voting protocols,
for one election with $n$ voters. }

\label{tab:compare-perf}
\begin{tabular*}{\columnwidth}{@{\extracolsep{\fill}}l|c|c|c|c|c@{}}

& \multicolumn{5}{c}{{\bf Protocol}} \\ 
 {\bf Metric} & KY~\cite{kiayias2002self} & G~\cite{groth_2004} & HRZ~\cite{hao_2010}  & 
KSRH~\cite{k12f} * & \name  \\\hline
{\scriptsize OT sessions}       & 0         & 0                & 0              & 0          &             $n-1$         \\\hline
{\scriptsize Exponentiations} & $2n^2+2n$    & $4n$                & $2n$              & $2n  $        &              $3n$          \\\hline
{\scriptsize ZKPs }           & $2n^2+2n$    & $4n$                & $2n$              & $2n $         &              0            \\\hline
{\scriptsize  Broadcast rounds}     & 3         &$n+1$             & 2              &  3         &              5  \\[2pt]
\end{tabular*}

{\quad} {\scriptsize * without robustness }
\end{table}

BVOT has significantly better performance than~\cite{kiayias2002self} and\linebreak \cite{groth_2004}
in that~\cite{kiayias2002self} requires a quadratic number of exponentiations
and ZKPs, and~\cite{groth_2004} requires a linear number of broadcast rounds,
as a function of the number of voters.  
BVOT is roughly similar in performance to~\cite{hao_2010} and~\cite{k12f},
with ZKPs replaced with OTs.

BVOT supports multiple candidates, 
whereas~\cite{kiayias2002self,groth_2004,hao_2010,k12f} require extensions with
additional costs to support multiple candidates.
BVOT and~\cite{k12f} are fair, but~\cite{kiayias2002self,groth_2004,hao_2010} are not.
In Table~\ref{tab:compare-perf}, we used a non-robust version of~\cite{k12f} because
none of the other protocols are robust; the robust version of~\cite{k12f}
requires additional costs.

\section{Security Notes} \label{sec:security}

We discuss the security properties of BVOT, including 
dispute freeness, perfect ballot secrecy,
fairness, robustness, and coercion resistance.
First, we give definitions of receiver and sender security for OT,
and we explain implications of choosing the parameter~$\lambda$.
Table~\ref{tab:compare-prop} summarizes our security comparison of
BVOT with four selected other self-tallying boardroom voting systems.

\subsection{Security of Oblivious Transfer} 

Adapting definitions from Naor and Pinkas~\cite{naor2005compsec}, we
state definitions of receiver and sender security for OT.
Receiver security means that the sender does not learn the receiver's choice
of strings.

\begin{definition}\label{def:rec-sec}
\textit{Receiver security in OT.} An $\textrm{OT}^m_1$ provides receiver security if and only if, for any probabilistic\linebreak
polynomial-time sender $\mathcal{A}$ with $m$ 
strings $s_1, s_2, \dots s_m$,\linebreak
given any $1 \leq i < j \leq m$ where the receiver chose either $s_i$ or $s_j$,
$\mathcal{A}$ cannot distinguish whether the receiver chose $s_i$ or $s_j$.
\end{definition}

Sender security means that the receiver does not learn anything other than the 
string they chose.
Sender security is defined in terms of a comparison between 
the information the receiver learns in the ideal implementation of oblivious transfer 
and the information the receiver learns in the real implementation.

\begin{definition}\label{def:sen-sec}
\textit{Sender security in OT.} 
An $\textrm{OT}^{\lambda m}_1$ provides sender security if and only if,
for every probabilistic\linebreak
polynomial-time receiver $\mathcal{B}$, 
substituting $\mathcal{B}$ in the real implementation of the protocol, 
there exists a probabilistic\linebreak
polynomial-time machine $\mathcal{B^{\prime}}$  
for the receiver’s role in the ideal implementation
such that, for every sequence of strings $s_1, s_2, \dots s_{\lambda m}$ of the sender, 
the outputs of $\mathcal{B}$ and
$\mathcal{B^{\prime}}$ are computationally indistinguishable.
\end{definition}

Because secure OT protocols (Definitions~1 and 2) are secure against malicious senders and receivers~\cite{peikert2008framework,chou2015simplest,asharov2017more},\linebreak
BVOT sufficiently handles malicious senders and receivers in the OT protocol.

\subsection{Choosing the Number of Primes for a Candidate}\label{lambda} 

The parameter $\lambda$ specifies the number of primes associated with each candidate. 
Its purpose is to mitigate the threat that the distributor might try to reduce some candidate's votes and increase some other candidate's votes (without changing
the total number of votes)
by exploiting its knowledge of the mapping ${\mathcal M}$.
Choosing $\lambda$ such that $\lambda m / n > 1$ is sufficiently large protects
against this threat.  
With this choice of $\lambda$, the distributor does not know 
the voter's selected prime.
If the distributor reduces the votes for an unchosen prime, the distributor will be caught.

\subsection{Dispute Freeness} 

\name is dispute free for the following reasons.
All communications are public and authenticated, and the threshold feature of the
encryption scheme ensures that all votes are included in the tally.

Manipulating the vote counts without detection requires knowledge of 
at least two distinct primes chosen by voters (e.g., voting with $p^2_1p^{-1}_2$
rather than with $p_1$).
This knowledge is hidden from the voters through the OT and masking.

If two voters collude by sharing their masked primes, they could unmask their
primes (see Section~\ref{sec:gcd}).  Assuming that $\lambda$ is chosen appropriately, 
the colluding voters would risk detection if they tried to cheat by adding more than
one vote to one of their candidates and subtracting one vote from the other---a result they
could have achieved directly without cheating.

The following elements prevent the distributor from engaging in malfeasance without detection: there are $\lambda$ primes per candidate, so the distributor does not know which primes were chosen.
Furthermore, the distributor commits to his vote, mask, and the mapping ${\mathcal M}$.

\subsection{Perfect Ballot Secrecy} 

BVOT enjoys perfect ballot secrecy for the following reasons.
First, the OT hides each voter's candidate selection.
Second, each voter encrypts their ballot.
The following theorem can be proven:

\begin{theorem} Assuming that $\textrm{OT}^{\lambda m}_1$ provides receiver security and the Diffie-Hellman assumption holds in the underlying group $G$, 
for each $1 \leq i \leq n$, 
voter $V_i$'s ballot $p_{r_i} g^{s}e^{x_i}$ is
indistinguishable from a randomly chosen element in $G$.
\end{theorem}

\subsection{Fairness} 

The threshold property of the encryption scheme ensures that none of the voters 
can learn anything about the tally until all of the voters publish their decryption share and the distributor publishes the unmasking value. Each voter casts their vote before any voter publishes their
decryption share.
Thus, BVOT is fair.

\subsection{Robustness} 
\label{sec:robust}

As is true for protocols \cite{kiayias2002self,groth_2004,hao_2010}, 
BVOT is not robust as defined below.

\begin{itemize}
\item \textbf{Robustness} ensures that the protocol can complete, even if one or
more voters attempt to prevent the protocol from completing.
\end{itemize}

Due to the threshold feature of the encryption scheme,
any one voter can prevent anyone from computing the tally
by withholding their vote.  
More generally, Kiayias et al.~\cite{kiayias2002self} claimed that no
self-tallying voting system that provides ballot secrecy is robust.

\subsection{Coercion Resistance}
\label{sec:coerce}

As is true for protocols 
\cite{kiayias2002self,groth_2004,hao_2010,k12f},
BVOT is not coercion resistant~\cite{JCJ2010} as defined below.

\begin{itemize}
\item \textbf{Coercion resistance} guarantees that a voter cannot prove 
to the adversary that they followed the adversary's demands.
\end{itemize}

BVOT is not coercion resistant because a voter can prove how they voted by
releasing their masked vote and the random value used in the 
encryption scheme.

\begin{table}[ht!]
\centering
\caption{Security properties of selected self-tallying boardroom voting protocols.}
\label{tab:compare-prop}
\begin{tabular*}{\columnwidth}{@{\extracolsep{\fill}}l|c|c|c|c|c@{}}
& \multicolumn{5}{c}{{\bf Protocol}}\\
\textbf{{\bf Property}} & KY~\cite{kiayias2002self} & G~\cite{groth_2004} 
& HRZ~\cite{hao_2010} & KSRH~\cite{k12f} & \name \\\hline
Fair                & $\times$& $\times$     & $\times$ & $\checkmark$ &  $\checkmark$   \\\hline
Perfect ballot privacy  & $\checkmark$ & $\checkmark$ & $\checkmark$ & $\checkmark$ &  $\checkmark$  \\\hline
Self-tallying           & $\checkmark$& $\checkmark$ & $\checkmark$& $\checkmark$ &  $\checkmark$   \\\hline
Dispute-free        &$\checkmark$ & $\checkmark$ & $\checkmark$& $\checkmark$ &  $\checkmark$   \\\hline
Robust            & $\times$& $\times$     &$\times$ & $\checkmark$ &  $\times$   \\\hline
Coercion resistant     & $\times$& $\times$     & $\times$ & $\times$     &  $\times$   
\end{tabular*}
\end{table}

BVOT has strictly better security properties than~\cite{kiayias2002self,groth_2004,hao_2010}  
in that~\cite{kiayias2002self,groth_2004,hao_2010} are not fair.  
BVOT and~\cite{k12f} have similar security properties, except that~\cite{k12f} is robust.
None of the protocols in Table~\ref{tab:compare-prop} achieve the
challenging property of coercion resistance.

\subsection{Disruptive or Dishonest Voters} 

It is possible that one or more of the voters (including the distributor) do not faithfully follow the protocol, perhaps in an attempt to cheat or disrupt the election.
We discuss three examples of such behaviors.  In each case, the
disruptive or dishonest behavior can be detected, though the source
of the behavior cannot necessarily be determined.

First, the distributor could misbehave, for example, by giving all voters the same masked prime or a different prime from the one specified by the mapping. 
Before tallying the votes, the other voters could detect such behaviors 
by examining the revealed mapping and its commitment.  
It might, however, be impossible to determine whether 
the distributor misbehaved or the voter lied in reporting malfeasance.

Second, one or more voters might generate $e$, encrypt, or decrypt incorrectly.
Anyone could detect such behaviors by attempting to compute the tally.

Third, instead of voting, for example, with a prime $p_1$, a voter could attempt to cheat
by voting with $p_1^2 p_2^{-1}$, 
where the voter is uncertain whether anyone else voted with $p_2 \neq p_1$.
If no one voted for $p_2$, then anyone could detect this behavior by
noticing that the tally includes a vote of $-1$ for $p_2$.
It might, however, be impossible to determine which voter caused this anomaly.

As explained in Section~\ref{sec:adversary}, in the spirit of boardroom
elections, we assume that voters will not intentionally engage in such
behaviors for the purpose of disrupting the election.

\section{Discussion}
\label{sec:discussion}

We discuss selected major design decisions, 
explain a variation of BVOT that uses an EA but is still self-tallying, 
point out an observation about masked primes,
and list some open problems.

\subsection{Major Design Decisions}\label{sec:design-dec} 

Two of our major design decisions were the following.
(1)~Instead of relying on each voter to engage in a complex ZKP of vote correctness, 
we use OT to hide each voter's ballot choice and to provide the voter with
the minimum information needed to cast their ballot.
(2)~We use multiparty threshold homomorphic encryption so that 
computing the vote tally is possible only when all votes are included.

\subsection{BVOT with Election Authority}\label{sec:EA} 

BVOT can also be used with a non-voting EA, in which the EA performs the administrative steps
of the distributor.  If the EA is untrusted (and might exfiltrate secrets), then
$\lambda$ must be chosen as before.  If the EA is trusted not to exfiltrate secrets, then
$\lambda = 1$ may be chosen.  With $\lambda=1$, BVOT runs faster, 
performing $n$ sessions of $\textrm{OT}^{m}_1$ 
instead of $n$ sessions of $\textrm{OT}^{\lambda m}_1$.

\subsection{An Observation about Masked Primes} 
\label{sec:gcd}

It is important that each voter does not learn more than one masked prime.
Otherwise, they could determine the mask value $g^s$ by computing 
$\gcd(p_ig^s, p_jg^s)$, where $p_ig^s$ and $p_jg^s$ are two distinct masked primes (Section~\ref{sec:vote-prt}).

Therefore, one cannot directly substitute a {\it private information retrieval (PIR)}~\cite{chor95pir} protocol for OT
in BVOT.  A straightforward such substitution would not prevent the voter from possibly learning
more than one masked prime.

\subsection{Open Problems} 

Open problems include:
(1)~Investigate the usability of BVOT.
(2)~Explore BVOT's possible use (with or without an EA) in particular applications, including 
as a consensus protocol for a distributed ledger system---for example, Hyperledger Fabric~\cite{hlf}.
(3)~Design remote boardroom voting protocols that are 
accountable (no one can make a false claim of malfeasance without detection) and coercion resistant.

\section{Conclusion}
\label{sec:conclude}

We introduced BVOT, a self-tallying boardroom voting protocol with 
fairness, perfect ballot secrecy, and dispute-\linebreak freeness. 
\name is the first boardroom voting protocol to use oblivious transfer to provide these properties.  
Unlike some existing protocols, \name avoids the complex
steps of requiring voters to carry out and check ZKPs.
BVOT illustrates the power and flexibility of oblivious transfer 
as a building block in protocol design.

\section*{Acknowledgments}
\label{sec:Acks}

We thank Enka Blanchard, Feng Hao, and Jonathan Katz for helpful comments.
Sherman was supported in part by
the National Science Foundation under SFS grant DGE-1753681, 
and by the U.S. Department of Defense under 
CySP grant H98230-19-1-0308.

\bibliographystyle{elsarticle-num}
\bibliography{short-list}

\begin{thebibliography}{10}
\expandafter\ifx\csname url\endcsname\relax
  \def\url#1{\texttt{#1}}\fi
\expandafter\ifx\csname urlprefix\endcsname\relax\def\urlprefix{URL }\fi
\expandafter\ifx\csname href\endcsname\relax
  \def\href#1#2{#2} \def\path#1{#1}\fi

\bibitem{blanchard2020boardroom}
E.~Blanchard, T.~Selker, A.~T. Sherman, Boardroom voting: Verifiable voting
  with ballot privacy using low-tech cryptography in a single room (2020).
\newblock \href {http://arxiv.org/abs/2007.14916} {\path{arXiv:2007.14916}}.

\bibitem{kiayias2002self}
A.~Kiayias, M.~Yung, Self-tallying elections and perfect ballot secrecy, in:
  International Workshop on Public Key Cryptography, Springer, 2002, pp.
  141--158.

\bibitem{groth_2004}
J.~Groth, Efficient maximal privacy in boardroom voting and anonymous
  broadcast, in: International Conference on Financial Cryptography, Springer,
  2004, pp. 90--104.

\bibitem{hao_2010}
F.~Hao, P.~Y. Ryan, P.~Zieli{\'n}ski, Anonymous voting by two-round public
  discussion, IET Information Security 4~(2) (2010) 62--67.

\bibitem{k12f}
D.~Khader, B.~Smyth, P.~Ryan, F.~Hao, A fair and robust voting system by
  broadcast, Lecture Notes in Informatics (LNI), Proceedings-Series of the
  Gesellschaft fur Informatik (GI) (2012) 285--299.

\bibitem{fujioka1992practical}
A.~Fujioka, T.~Okamoto, K.~Ohta, A practical secret voting scheme for large
  scale elections, in: International Workshop on the Theory and Application of
  Cryptographic Techniques, Springer, 1992, pp. 244--251.

\bibitem{Adida2008HeliosWO}
B.~Adida, Helios: Web-based open-audit voting, in: USENIX Security Symposium,
  Vol.~17, 2008, pp. 335--348.

\bibitem{clarkson2008civitas}
M.~R. Clarkson, S.~Chong, A.~C. Myers, Civitas: Toward a secure voting system,
  in: 2008 IEEE Symposium on Security and Privacy (sp 2008), IEEE, 2008, pp.
  354--368.

\bibitem{chaum2009scan}
D.~{Chaum}, R.~T. {Carback}, J.~{Clark}, A.~{Essex}, S.~{Popoveniuc}, R.~L.
  {Rivest}, P.~Y.~A. {Ryan}, E.~{Shen}, A.~T. {Sherman}, P.~L. {Vora},
  Scantegrity {II}: End-to-end verifiability by voters of optical scan
  elections through confirmation codes, IEEE Transactions on Information
  Forensics and Security 4~(4) (2009) 611--627.

\bibitem{remotegrity13}
F.~Zag{\'o}rski, R.~T. Carback, D.~Chaum, J.~Clark, A.~Essex, P.~L. Vora,
  Remotegrity: Design and use of an end-to-end verifiable remote voting system,
  in: International Conference on Applied Cryptography and Network Security,
  Springer, 2013, pp. 441--457.

\bibitem{hao2014every}
F.~Hao, M.~N. Kreeger, B.~Randell, D.~Clarke, S.~F. Shahandashti, P.~H.-J. Lee,
  Every vote counts: Ensuring integrity in large-scale electronic voting, in:
  2014 Electronic Voting Technology Workshop/Workshop on Trustworthy Elections
  (EVT/WOTE 14), 2014, pp. 1--25.

\bibitem{shahandashti2016dre}
S.~F. Shahandashti, F.~Hao, {DRE}-ip: a verifiable e-voting scheme without
  tallying authorities, in: European Symposium on Research in Computer
  Security, 2016, pp. 223--240.

\bibitem{kil88}
J.~Kilian, Founding cryptography on oblivious transfer, in: Proceedings of the
  twentieth annual ACM symposium on Theory of computing, 1988, pp. 20--31.

\bibitem{nurmi91}
H.~Nurmi, A.~Salomaa, L.~Santean, Secret ballot elections in computer networks,
  Computers \& Security 10~(6) (1991) 553--560.

\bibitem{Rabin81}
M.~O. Rabin, How to exchange secrets with oblivious transfer, Technical Report
  TR-81, Aiken Computation Lab, Harvard University (1981).

\bibitem{impagliazzo1989limits}
R.~Impagliazzo, S.~Rudich, Limits on the provable consequences of one-way
  permutations, in: Proceedings of the Twenty-First Annual ACM Symposium on
  Theory of Computing, STOC ’89, Association for Computing Machinery, New
  York, NY, USA, 1989, p. 44–61.

\bibitem{Bellare89}
M.~Bellare, S.~Micali, Non-interactive oblivious transfer and applications, in:
  Conference on the Theory and Application of Cryptology, 1989, pp. 547--557.

\bibitem{Naor01}
M.~Naor, B.~Pinkas, Efficient oblivious transfer protocols, in: 12th annual
  ACM-SIAM symposium on Discrete algorithms, 2001, pp. 448--457.

\bibitem{naor2005compsec}
M.~Naor, B.~Pinkas, Computationally secure oblivious transfer, Journal of
  Cryptology 18~(1) (2005) 1--35.

\bibitem{peikert2008framework}
C.~Peikert, V.~Vaikuntanathan, B.~Waters, A framework for efficient and
  composable oblivious transfer, in: Annual international cryptology
  conference, Springer, 2008, pp. 554--571.

\bibitem{chou2015simplest}
T.~Chou, C.~Orlandi, The simplest protocol for oblivious transfer, in:
  International Conference on Cryptology and Information Security in Latin
  America, Springer, 2015, pp. 40--58.

\bibitem{asharov2017more}
G.~Asharov, Y.~Lindell, T.~Schneider, M.~Zohner, More efficient oblivious
  transfer extensions, Journal of Cryptology 30~(3) (2017) 805--858.

\bibitem{stinson2018cryptography}
D.~R. Stinson, M.~Paterson, Cryptography: {T}heory and {P}ractice, CRC press,
  2018.

\bibitem{s15n}
A.~Szepieniec, B.~Preneel, New techniques for electronic voting, {USENIX}
  Journal of Election Technology and Systems ({JETS}) 3~(2) (2015) 46--69.

\bibitem{giustolisi2016possibility}
R.~Giustolisi, V.~Iovino, P.~B. R{\o}nne, On the possibility of non-interactive
  e-voting in the public-key setting, in: International Conference on Financial
  Cryptography and Data Security, Springer, 2016, pp. 193--208.

\bibitem{cramer97}
R.~Cramer, R.~Gennaro, B.~Schoenmakers, A secure and optimally efficient
  multi-authority election scheme, European transactions on Telecommunications
  8~(5) (1997) 481--490.

\bibitem{crepeau1995committed}
C.~Cr{\'e}peau, J.~van~de Graaf, A.~Tapp, Committed oblivious transfer and
  private multi-party computation, in: Annual International Cryptology
  Conference, Springer, 1995, pp. 110--123.

\bibitem{even85}
S.~Even, O.~Goldreich, A.~Lempel, A randomized protocol for signing contracts,
  Communications of the ACM 28~(6) (1985) 637--647.

\bibitem{fagin96}
R.~Fagin, M.~Naor, P.~Winkler, Comparing information without leaking it,
  Communications of the ACM 39~(5) (1996) 77--85.

\bibitem{benaloh2006}
J.~Benaloh, Simple verifiable elections, in: Proceedings of the USENIX/Accurate
  Electronic Voting Technology Workshop 2006 on Electronic Voting Technology
  Workshop, EVT’06, USENIX Association, USA, 2006, p.~5.

\bibitem{JCJ2010}
A.~Juels, D.~Catalano, M.~Jakobsson, Coercion-resistant electronic elections,
  in: Towards Trustworthy Elections, Springer, 2010, pp. 37--63.

\bibitem{chor95pir}
B.~Chor, O.~Goldreich, E.~Kushilevitz, M.~Sudan, Private information retrieval,
  in: Proceedings of IEEE 36th Annual Foundations of Computer Science, IEEE,
  1995, pp. 41--50.

\bibitem{hlf}
E.~Androulaki, A.~Barger, V.~Bortnikov, C.~Cachin, K.~Christidis, A.~De~Caro,
  D.~Enyeart, C.~Ferris, G.~Laventman, Y.~Manevich, et~al., Hyperledger fabric:
  a distributed operating system for permissioned blockchains, in: Proceedings
  of the thirteenth EuroSys conference, 2018, pp. 1--15.

\end{thebibliography}

\end{document}